\documentstyle[12pt]{article}

\newcommand{\al}{\alpha} 
 
\newcommand{\als}{\frac{\alpha_s}{\pi}}

\newcommand{\ice}[1]{\relax}
\newcommand{\be}{\begin{equation}}
\newcommand{\ee}{\end{equation}}
\newcommand{\MSsch}{\overline{\rm MS}}
\begin{document} 
\begin{flushright}
{\bf MZ-TH/99-37}\\
\end{flushright}
\vspace*{0.4cm}

\begin{center} 
{\Large \bf Asymptotic structure of perturbative series
for $\tau$ lepton observables
}
\vspace{0.6truecm}

{\large J.G.~K\"orner$^1$, F.Krajewski$^1$, A.A.~Pivovarov$^{1,2}$}\\[.1cm]
$^1$ Institut f\"ur Physik, Johannes-Gutenberg-Universit\"at,\\[-.1truecm]
  Staudinger Weg 7, D-55099 Mainz, Germany\\[.2truecm]
$^2$ Institute for Nuclear Research of the\\[-.1truecm]
  Russian Academy of Sciences, Moscow 117312
\vspace{0.6truecm}
\end{center}

\vskip 1cm
\centerline {\bf Abstract}
\noindent We analyze $\tau$ lepton decay observables, namely
moments of the hadronic spectral density in the finite energy interval
$(0,M_\tau)$, within
finite order perturbation theory including $\al_s^4$ corrections.
The start of the asymptotic growth of the perturbation theory series
is found at this order in a scheme invariant manner.
We establish the ultimate accuracy of finite order
perturbation theory predictions and 
discuss the construction of optimal observables.
\thispagestyle{empty}
\newpage
\section{Introduction}
The study of $\tau$ lepton decays provides a wealth of information on low energy
hadronic physics where the accuracy
of experimental data is permanently improving
\cite{exp,PDG}. 
The central quantity of interest is 
the hadronic spectral density. The spectral density
has been calculated with a very high
degree of accuracy within perturbation theory (e.g. \cite{phys_report,eek2,eek2c}). 
The structure of observables -- related to the two-point correlator of
hadronic currents with well established and simple analytic properties
-- makes the comparison of experimental data with
theoretical calculations very clean. All these features make
$\tau$ lepton physics an important area of particle phenomenology 
where theory (QCD) can be confronted with experiment 
to a very high precision
\cite{SchTra84,Bra88,Bra89,NarPic88,BraNarPic92}.

In the present note we show
that within the finite order
perturbation theory analysis the ultimate
theoretical precision has been reached already now.
The limit of precision exists due to the asymptotic nature
of the perturbation theory series. 
The actual magnitude of this limiting precision 
depends on 
the numerical value of the coupling constant which is the expansion
parameter.
We perform our analysis and reach 
our conclusions in a renormalization scheme invariant way.

The normalized $\tau$ lepton decay rate into hadrons $h$ is given by 
\be 
  \label{rate}
R_\tau={\Gamma(\tau \rightarrow h\nu)\over 
\Gamma(\tau \rightarrow l\nu\bar\nu)}
=N_c(1+\delta)
\ee
with 
\be
\label{expdec}
R_\tau^{exp}=3.649\pm0.014 \quad {\rm  and}\quad 
\delta^{exp}=0.216\pm0.005 \ .
\ee
The first term in eq.~(\ref{rate}) is the parton model 
result while the second term 
$\delta$ represents the effects of QCD interaction.
In this paper we neglect electroweak corrections altogether. 
The theoretical expression for the rate is given by 
\begin{equation}
  \label{int}
R_\tau=N_c\int_0^{M_\tau^2}2(1-{s\over M_\tau^2})^2(1+2{s\over M_\tau^2})
\rho(s){ds\over M_\tau^2} \ .
\end{equation}
The spectral density $\rho(s)$ is related to Adler's
$D$-function through the dispersion relation 
\be
  \label{disp}
D(Q^2)=Q^2\int {\rho(s)ds \over (s+Q^2)^2} \ .
\end{equation}
The $D$-function is computable in perturbation theory. 
In the $\overline{\rm MS}$ scheme the perturbation theory expression for the 
$D$-function is given by 
\begin{equation}
  \label{spect}
D(Q^2)=1+\als+1.64 \left(\als\right)^2 + 6.37  \left(\als\right)^3 + k_3
 \left(\als\right)^4
+\ldots 
\end{equation}
where the running coupling is normalized at the scale 
$\mu=Q$. The light quarks $u$, $d$ and $s$ are taken to be massless. 
Eqs.~(\ref{int}-\ref{spect})
constitute 
the full theoretical information 
necessary for the perturbation theory analysis of
the $\tau$ system. The fourth order 
$\overline{\rm MS}$-scheme coefficient $k_3$ is not known
at present.

In the present note we do not systematically 
discuss non-perturbative effects stemming 
from standard power corrections \cite{SVZ}.
Also, the infinite resummation of the perturbation theory series
different from the standard renormalization group improvement
is used only as a toy example \cite{RG}. 
The standard power corrections due to nonvanishing vacuum expectation
values
of local operators 
within operator product expansion are relatively small and can be simply
accounted for if desired. The coefficient
functions of local operators are 
known in low orders of the perturbation theory expansions
and 
there is no necessity to thoroughly analyze their convergence
properties.
It is the high precision achieved in the experimental
analysis of $\tau$ decays
and the rather advanced stage of theoretical description
that calls for a detailed analysis of
the physics of the $\tau$ system.

\section{Internal perturbation theory description of basic $\tau$
system observables}
The central quantity of interest in the $\tau$ system
is the hadronic spectral density which can be measured in the 
finite energy interval $(0,M_\tau=1.777~{\rm GeV})$.
Being a distribution (in theory)
or a rapidly varying function in the vicinity of resonances
(in experiment)
the hadronic spectral density cannot be analyzed 
pointwise within perturbation theory.
The appropriate quantities to be analyzed are the moments (or generalized
Fourier components over a chosen complete set of test functions).
We define moments of the spectral density by
(with $M_\tau$ chosen to be the unity of mass)
\begin{equation}
  \label{intmom}
M_n=(n+1)\int_0^1 \rho(s) s^n ds \equiv 1+m_n  \ .
\end{equation}
Due to the completeness of the basis $\{s^n:n=0,\ldots,\infty\}$ 
the moments $m_n$
contain the entire
information about $\rho(s)$. 
The invariant
content of the investigation of the spectrum,
i.e. independent of any definition of the charge,
is the simultaneous analysis of all the moments.
Note that within finite order
perturbation theory the moments eq.~(\ref{intmom})
coincide with the results of contour integration
\cite{cont,cont1,cont2,Pivtau}
because of analytic properties of the functions $\ln^p s$.

In order to get rid of artificial scheme-dependent constants in 
the perturbation theory expressions for the moments
we define an effective coupling $a(s)$ directly on the physical cut 
through the relation
\begin{equation}
  \label{defofa}
\rho(s)=1+a(s) \ .
\end{equation}
All the constants that may appear due to a particular
choice of the renormalization 
scheme 
are absorbed into the definition of the effective charge e.g.
\cite{effsch,ksch,kksch,effDh}.
Note that if there was no running (as in the conformal limit of QCD
with vanishing $\beta$-function or at the infrared fixed point)
then the whole physics of the $\tau$ system in the massless approximation
(without strange particles, for instance, and including only 
perturbative corrections
without possible power corrections) 
would reduce to the determination of a single number 
$a(M_\tau)\equiv a$ and consequently 
there would not be any problems with the 
convergence of the perturbation theory series.
Because of the running of $a(s)$, however, different observables, i.e.
different moments of the spectral density, generate different 
perturbation theory
series from the original object $\rho(s)$ in eq.~(\ref{defofa}).
The whole set of moments needs to be
analyzed in a scheme invariant way e.g. 
\cite{prl,brodsky1,brodsky}. Note that the introduction of
a natural internal coupling parameter 
such as the effective charge $a(s)$, 
allows one to extend 
the perturbation theory series needed for the description of relations
between observables
by one more term
as compared to the analysis in e.g. the $\MSsch$ scheme (e.g. \cite{prl,renRS}).
When defining the effective charge directly through $\rho(s)$ itself 
we get theoretical perturbative corrections to the moments only
because of running. Without running one would have 
\be
\label{norunmom}
M_n=1+a(M_\tau)\equiv 1+a\quad {\rm or} \quad   m_n\equiv a
\ee
and the perturbation theory analysis would be over
(we neglect power corrections for the moment!). 
In any given order of perturbation theory the
running of the coupling $a(s)$ defined in eq.~(\ref{defofa})
contains only logarithms of $s$ with coefficients given by an effective
$\beta$-function
\be
\label{run}
a(s) = a + \beta_0 L a^2 + (\beta_1 L + \beta_0^2 L^2) a^3
+ (\beta_2 L +\frac{5}{2}\beta_1\beta_0 L^2 + \beta_0^3 L^3) a^4 + \ldots
\ee
where $a=a(M_\tau^2)$, $L=\ln(M_\tau^2/s)$.
The contributions of powers 
of logarithms to the normalized moments are 
\begin{equation}
\label{logs}
(n+1)\int_0^1 s^n \ln^p(1/s)ds = \frac{p!}{(n+1)^p} \ .
\ee
Therefore, at fixed order of 
perturbation theory the effects of running die out for
large $n$ improving the convergence of perturbation theory series. 
With the definition of the charge according to eq.~(\ref{defofa})
all high order corrections vanish at $n\rightarrow \infty$ at any 
fixed order of perturbation theory. 
With running one has instead of eq.~(\ref{norunmom}) 
\begin{eqnarray}
\label{momff}
m_0&=&a + 2.25 a^2  + 14.13 a^3  + 87.66 a^4 
+ (433.3 + 4.5 k_3) a^5   \nonumber \\
m_1&=&a + 1.125 a^2  + 4.531 a^3  + 6.949 a^4 + (-175.2 + 2.25
k_3) a^5   \nonumber \\
m_2&=&a + 0.75 a^2  + 2.458 a^3  - 1.032 a^4  + 
(-142.6 + 1.5 k_3) a^5 \nonumber \\
m_3&=&a + 0.563 a^2  + 1.633 a^3  - 2.542 a^4  + 
(-110.4 + 1.125 k_3) a^5\nonumber \\
&&\cdots \nonumber \\
m_{100}&=&a + 0.022 a^2  + 0.041 a^3  - 0.25 a^4  + 
(-4.08 + 0.045 k_3) a^5
\end{eqnarray}
For large $n$ the moments behave better because the infra-red region 
of integration is suppressed.
Note that the coefficients of the series in 
eq.~(\ref{momff}) are saturated with the lowest power of logarithm
for large $n$ for a given order of perturbation theory, i.e. 
they are saturated with the highest coefficient of 
the effective $\beta$-function.

Higher moments are not welcome from the experimental point of view.
They are dominated by the contributions coming from the 
high energy end of the $\tau$ decay spectrum (therefore they converge
better perturbatively) but
experimental accuracy for the moments basically deteriorates with
increasing $n$ because poorly known
contributions close to the right end of the interval are enhanced.
To suppress experimental errors
from the high energy end of the spectrum
the modified system of mixed moments  
\begin{equation}
  \label{intmomkl}
\tilde M_{kl}=\frac{(k+l+1)!}{k!l!}\int_0^1 \rho(s) (1-s)^k s^l ds
\equiv 1+\tilde m_{kl}
\end{equation}
can be used e.g. \cite{DP}.
The weight function $(1-s)^k s^l$ has its maximum at $\bar s =l/(l+k)$.
The integral in eq.~(\ref{intmomkl}) is dominated 
by contributions from around this value.
A disadvantage of choosing such moments
is that the $(1-s)^k$ factor enhances the infra-red region strongly and
ruins the perturbation theory convergence.
As an example one has
\begin{eqnarray}
\label{altmom}
\tilde m_{00}&=&a + 2.25 a^2  + 14.13 a^3  + 87.66 a^4 + (433.3 + 4.5
k_3) a^5   \nonumber \\
\tilde m_{10}&=&a + 3.375 a^2  + 23.72 a^3  + 168.4 a^4  + 
 (1042. + 6.75 k_3) a^5 \nonumber \\ 
\tilde m_{20}&=&a + 4.125 a^2  + 31.24 a^3  + 241.1 a^4  + 
(1683. + 8.25 k_3) a^5 \nonumber \\ 
\tilde m_{30}&=&a + 4.688 a^2  + 37.51 a^3  + 307.3 a^4  + 
 (2324. + 9.375 k_3) a^5 \ .
\end{eqnarray}
The values of the coefficients in the series eq.~(\ref{altmom})
can be found in a concise form for arbitrary
$k$ at any giving finite order of perturbation theory. 
For instance, the contribution of the $log$-term is given by
\be
\label{logalt}
(k+1)\int_0^1 (1-s)^k \ln(1/s) ds = \sum_{j=1}^{k+1} \frac{1}{j} \ .
\end{equation}
In contrast to eq.~(\ref{logs}) it increases as $\ln(k)$ for large
$k$ making the coefficients of the perturbation theory series large.
The contribution of $log^2$-term reads
\be
\label{logalt2}
(k+1)\int_0^1 (1-s)^k \ln^2(1/s) ds 
= \left(\sum_{j=1}^{k+1} \frac{1}{j} \right)^2
+ \sum_{j=1}^{k+1} \frac{1}{j^2} 
\end{equation}
and can be seen to grow as $\ln^2(k)$ for large $k$.

In practical applications
our formal criterion of the accuracy which the series provides
is given by the numerical magnitude of the last term of the series.
However, this criterion should be applied with great caution. 
Because of the freedom of
the redefinition of the expansion parameter the last term of the
series
can always
be made arbitrary small for any given observable. 
One can give an invariant meaning to 
the quality of the perturbation theory expansion only for a set of
observables.

Before proceeding we would like to 
comment on the contribution of power corrections
to the systems of moments eqs.~(\ref{intmom},\ref{intmomkl}) and 
the interplay
between the magnitude of this contribution 
and the structure of the perturbation theory series.
For the system of moments in eq.~(\ref{intmom}) 
the contribution of power corrections 
reduces to a single term (neglecting the weak
$log(Q)$ dependence of
coefficient functions of the operator product expansion
which is a common practice)
of the form $(\Lambda^2/M_\tau^2)^n$ which decreases
very fast with $n$
($\Lambda$ is a typical scale of
power corrections related to the non-perturbative scale of QCD 
and $\Lambda < M_\tau$).
This makes the perturbation theory contribution dominant 
in the total result.
This perturbative term for the large $n$ moments
is saturated with high energy contributions and therefore
converges perturbatively. 
The
convergence becomes even better with increasing $n$. 
The moments are perturbatively dominated and, therefore, 
precise.  
On the contrary, for 
the system of mixed moments in eq.~(\ref{intmomkl}) with $l\sim 0$
the large $k$ moments
are saturated with low energy contributions, i.e. basically with the
contribution of the ground state resonance, and therefore
are completely nonperturbative which is reflected in 
the fast deterioration of perturbative convergence. 
The contribution of power corrections to the moments
eq.~(\ref{intmomkl})
picks out many terms in all orders from $n=2$ to $n=k+1$.
Nothing definite can be said about such a sum of power corrections
in any realistic case.
This also indicates the importance of power corrections
for mixed moments.
The perturbation theory series is 
the same both for the vector and axial vector
channels while 
the lowest resonance contributions are completely different 
(the pion instead of the $\rho$-meson). 
Therefore no method of summation of perturbation theory series can bring 
it to agree with experiment. In this case perturbation theory is in trouble and
the power corrections provide the correct result for large $k$.
Large $k$ mixed moments, therefore, are not usable within the 
perturbation theory framework
even if they are preferable from 
the experimental point of view.

Thus, one faces the usual clash between experimental and
theoretical accuracy 
which is reflected in our case in 
the range of $(k,l)$ values for the 
mixed moments that are chosen as optimal observables.
Having explicit perturbation theory 
formulas at hand one can establish the ultimate 
theoretical accuracy implied by 
the asymptotic character of perturbation theory series
for a given experimental observable with any stated
precision. This allows one to conclude which
error
--  experimental or theoretical -- dominates the uncertainty of 
an observable related to $\tau$ decay physics.

For our numerical estimates we take $a=0.111$ as
obtained
from the corresponding value of the $\MSsch$ charge.
From the set of moments $\{m_n;n=0,\ldots,\}$
the moment $m_0$ has the biggest infra-red contribution. 
Therefore a set of observables
has the worst accuracy if the moment $m_0$ is included in the set.
For $m_0$ one obtains
\be
\label{num0}
m_0=0.111 + 0.0277 + 0.0193 + 0.0133 + 
(0.0073 + 0.000076 k_3) \ .
\ee
As mentioned before
the numerical value of the coefficient $k_3$
is unknown at present.
In some of the following evaluations we want to fix its value 
to have a feeling of the importance of the last term
of the perturbation theory series.
One popular value is 
$k_3=25$ based on Pad\'e approximation.
Another value $k_3=91$
nullifies the fourth order coefficient of 
the effective $\beta$-
function \cite{groote}.
Both these numbers are used only for illustrative purposes while 
our conclusion about perturbatively commensurate observables is
independent of $k_3$.
Numerically for $k_3=25$
one has for the zeroth moment
\be
\label{num025}
m_0=0.111 + 0.0277  + 0.0193   + 0.0133   +  ``0.009'' \ .
\ee
Formally, the convergence still persists even in eq.~(\ref{num025}) 
if one only requires subsequent terms
of the series to 
decrease but the convergence is very slow. 
Also, the total contribution of the four higher order terms is more
than 
60\% 
of the leading one.
For the first moment the convergence is considerably better
\be
\label{num1} 
m_1= 0.111 + 0.014 + 0.006 + 0.001+(-0.003 + 0.00004 k_3) 
\ee
and for $k_3=25$
\be
\label{num125}
m_1= 0.111 + 0.014  + 0.006  + 0.001 - ``0.002'' \ .
\ee
With the choice $k_3=25$
the $O(a^5)$ term already shows numerical growth.
If one keeps only the smallest term 
one gets a formal accuracy of
about 1\% and the total contribution of the three higher order 
terms gives only about 20\% of the leading one.
The large difference in accuracy between $m_0$ and $m_1$
is a general feature of the moment observables at fifth order of perturbation theory:
one cannot get a uniform
smallness at this order for several moments at the same time 
adjusting only one number
$k_3$.
For $a=0.111$ we therefore conclude that one is starting
to see the onset of asymptotic growth at fifth
order.
The growth of the terms is independent of any definition of
the charge if several moments are analyzed simultaneously and this
feature 
cannot be changed by any choice of $k_3$.
For any single moment, e.g. $m_0$,
one can always redefine the charge and make the
series converge well at any desired rate 
but then other moments become bad in terms of this charge.
The invariant statement about the asymptotic
growth is that the system of moments $m_n$ with $n=0$
included cannot be treated perturbatively at the fifth order 
of perturbation theory 
for the numerical value of the expansion parameter 
$a=0.111$ if one wants to obtain an accuracy better
than 5\% - 10\%. This statement about the ultimate
accuracy of the set of moment observables
attainable in 
fifth order of perturbation theory is independent of whichever numerical 
value $k_3$ takes. 
If, however, the system of moments $m_n$
does not include $m_0$ as an observable,
a uniform accuracy better than 1\% can be obtained for 
such a system within perturbation theory.
For instance, excluding $m_0$ and using 
$k_3 \sim 100$ one can make the system of moments with $n\ge 1$
perturbation theory commensurate at fifth order in a sense that all fifth order
terms can be made small simultaneously.
To demonstrate this in a scheme invariant way we choose the second
moment (which is already well convergent) as a definition of our
experimental charge and find 
\begin{eqnarray}
\label{invmom}
m_{0}&=&m_2 + 1.5 m_2^2 +  9.417 m_2^3 +  59.28 m_2^4+(310.3 +
3 k_3) m_2^5 
\nonumber \\
m_{1}&=&m_2 + 0.375 m_2^2 +  1.51 m_2^3 + 2.527 m_2^4
+(-54.45 + 0.75 k_3) m_2^5 \nonumber \\ 
m_{2}&=&m_2 \nonumber \\ 
m_{3}&=&m_2 - 0.19 m_2^2 - 0.544 m_2^3 + 0.742 m_2^4+(35.2 -
0.375 k_3) m_2^5
\nonumber \\ 
m_{4}&=&m_2 - 0.3 m_2^2  - 0.803 m_2^3  + 1.69 m_2^4 +
(56.641 - 0.6 k_3) m_2^5  \ .
\end{eqnarray}
The convergence for the moments $m_1$-$m_4$ 
(and for $n>4$) is fine. 
The total contribution
of higher order corrections is small. 
The worst series is the one for
the zeroth order moment. Eq.~(\ref{invmom})
shows that no choice of $k_3$ yields
accuracy for both $m_0$ and $m_1$ which is essentially 
better than the fourth order
term.
In fact, there is a narrow window $40<k_3<60$
where the formal criterion of convergence is satisfied for both $m_0$
and $m_1$ but we do not find it natural to rely on such a fine tuning
and even then the accuracy of the zero moment is only about 10\%. 
This is an indication that the ultimate 
accuracy of the perturbation theory expansion
for the zeroth moment 
has been reached.
If the moment $m_0$ is excluded the choice $k_3\sim 100$
allows one to make the convergence fast 
even to fifth order and no
conclusion about an asymptotic growth is possible.

The perturbation theory expansions for 
the system of moments with $(1-s)^k$ weight shows worse behavior.
With the above criterion of accuracy,
the precision which is given by the series from eq.~(\ref{altmom})
is of order 10\% - 20\% for the numerical value of $a$. 
This is not enough for a comparison with
experiment at the present level of precision.
For instance, an expansion of the higher moments in eq.~(\ref{altmom})
in terms of the first one 
(which is the most perturbative one for this system) goes as follows
\begin{eqnarray}
\label{altmomnum}
\tilde m_{00}&\equiv&m_0 = 0.17 \nonumber \\
\tilde m_{10}&=&0.17  
+ 0.033 + 0.022 + 0.011 + (-0.005 + 0.00032 k_3)\nonumber \\ 
\tilde m_{20}&=&0.17 
+ 0.054 + 0.043 + 0.027 + ( 0.0015+ 0.00053 k_3)\nonumber \\
\tilde{m}_{30} &=& 0.17 + 0.070 + 0.061 + 0.046 + ( 0.014+ 0.0007 k_3 )
\end{eqnarray}
These series possess a formal accuracy of from 6\% to 25\%
and the contribution of higher order terms can be as large as 
the leading term.
Because of the slow convergence
there is no sign of improvement with higher order of the perturbation theory:
the series expansions do not allow 
any reliable estimate of accuracy for large mixed moments.
Also while for the moments eq.~(\ref{intmom})
the total contribution of
corrections is small, the situation is different here.
The total change of the leading order result 
due to higher order corrections is considerable
and strongly differs for various moments.
This is another indication that the set of mixed moments is not 
commensurate perturbatively. 

\section{$\tau$ decay rate}
The $\tau$ decay width is given by a specific linear
combination of moments.
Because of the factor $(1-s)^2$ present in eq.~(\ref{int})
the convergence property of the total decay rate observable is
not optimal.
The $(1-s)^2$ factor enhances the infra-red region of integration, i.e. 
the relative magnitude
of the contributions of logarithms $\ln(M_\tau^2/s)$
at small energy.
The concrete shape of the weight function with 
the weight factor $(1-s)^2$ is the main source of slow convergence.
One has 
\be
\label{rtau}
r_\tau=a + 3.563 a^2  + 24.97 a^3  + 
    174.8 a^4  + (1041. + 7.125 k_3) a^5 \ .
\ee
Using $a=0.111$ and $k_3=25$
\be
\label{rtau0111}
r_\tau = 0.111 + 0.044 + 0.034 + 0.027 + ``0.021'' 
\ee
Formally the consecutive terms decrease but the decrease is very slow.
One can see that the pattern of convergence mainly follows that of 
the moment $\tilde m_{20}$ from eq.~(\ref{altmom})
because of the factor $(1-s)^2$ in eq.~(\ref{int}).

Eqs. (\ref{expdec}) and (\ref{rtau0111})
show the essence of the problem we are addressing.
In the finite order perturbation theory analysis
one has to compare $\delta^{exp}$ with $\delta^{th}$ 
\be 
\label{theoac}
0.216\pm 0.005 = \delta^{exp} \quad {\rm vs} \quad
\delta^{th} = 0.111 + 0.044 + 0.034 + 0.027 + ``0.021''
\ee
and the uncertainty of the theoretical
expression ``0.021'' (or even 0.027) is much larger than the
experimental error 0.005. 
Thus, the theoretical uncertainty due to the truncation of 
the perturbation theory series
is much larger than the experimental 
error of the corresponding observable.
The common practical
expectation for theoretical perturbation theory expansions to be useful is 
the smallness of the total higher order corrections
if nothing is known about the convergence of the expansion. 
For the rate  
observable the corrections increase the leading order result
by a factor of 2. Note that one can improve the explicit 
convergence of the rate
observable by a special redefinition of the expansion
parameter due to renormalization scheme freedom.
However, then the first moment of the differential decay rate
will behave wildly. 
It is this feature that prompts us to
reach definite 
conclusions about the asymptotic growth of perturbation theory expansion 
independent of
any scheme. Two different sets of observables
where one set includes the moment $m_0$ and the other set does not
include 
it 
are not perturbatively connected
with an accuracy required by experiment.
Indeed, the first $s$-moment of 
the differential decay rate $d R_\tau/ds$ gives
the series with faster convergence than eq.~(\ref{rtau}) 
\be
\label{rtau1}
r_\tau^{(1)} =
a + 2.138 a^2  + 10.15 a^3  +  28.43 a^4  +
 (-268.3 + 4.275 k_3) a^5
\ee
or numerically with $k_3=25$ 
\be
\label{rtau10111}
r_\tau^{(1)} =
0.111 + 0.026 + 0.014 + 0.004 - ``0.003'' \ .  
\ee       
The second $s$-moment has even better perturbative expansion
\be
\label{rtau2an} 
r_\tau^{(2)}=a + 1.575 a^2  + 6.186 a^3  + 6.386 a^4 +
 (- 283.3  + 3.15 k_3)a^5
\ee
and numerically with $k_3=25$ 
\be
\label{rtau2}
r_\tau^{(2)}
=
0.111 + 0.0194 + 0.0085 + 0.001 - ``0.003'' \ .
\ee
The fifth order term is larger than the fourth order term for 
$k_3=25$.
No choice of $k_3$ can simultaneously
make all these three observables convergent at fifth
order.
If one chooses $k_3\sim 100$ in order 
to guarantee for a better convergence 
of the higher moments (which is 
physically motivated) one almost destroys
the perturbation theory series
for the decay rate eq.~(\ref{rtau}).

The $(1-s)^n$
moments of the differential decay rate
suppress poorly known high energy experimental data. 
Taking $n=1$ as an example one has
\be
\label{alttauan} 
r_\tau^{(1-s)}= a + 4.173 a^2  + 31.31a^3  + 237.6 a^4  + 
(1603.  + 8.35 k_3) a^5
\ee
and numerically for $k_3=25$
\be
\label{alttau}
r_\tau^{(1-s)}= 0.111 + 0.051 + 0.043   + 0.036 +  ``0.031'' \ .
\ee 
For $k_3=100$ the series reads
\be
\label{alttaunum}
r_\tau^{(1-s)} = 0.111+ 0.051 + 0.043 + 0.036 + ``0.041''
\ee 
which gives only about 30\% accuracy and more than a factor 2 change of
the leading order term. 
We conclude that the theoretical 
precision cannot compete with the experimental precision.

There are two distinct problems in analyzing 
$\tau$ decays: one is to describe the set of
observables of the system using its internal coupling parameter
defined to get the highest precision
and establish whether the set is perturbatively commensurate, 
while
another is 
to extract the standard $\MSsch$ 
parameters. It can happen that the set of observables  
is perturbatively connected with some given accuracy
but the $\MSsch$ coupling $\al_s$ is not the best parameter for the expansion.
This is the case here.
In internal terms the $\tau$ system is described with
higher accuracy in terms of the number of perturbation theory terms
than in the $\MSsch$ scheme. However, at this level of
expansion one sees the asymptotic growth of the perturbation series
for the numerical value of the expansion parameter fixed by experiment.

The expression for the decay rate
in the $\MSsch$ scheme possesses only
$O(\al_s^3)$ accuracy
\be
\label{taumssch}
r_\tau = \left(\als\right)+5.20 \left(\als\right)^2
+26.4\left(\als\right)^3
+(78.0+k_3)\left(\als\right)^4
\ee
with a numerical precision of only 30\% again.
A numerical value for $\al_s$ is
usually extracted treating
the three first terms of the rate expression eq.~(\ref{taumssch})
as an exact function. The numerical value found is rather precise.
However, the accuracy of the numerical prediction for
other observables is dominated by the uncertainty of truncation 
of the series and is poor if the observable contains the zeroth order
moment. Therefore 
the comparison of different observables of the system 
cannot be done with high precision
and the ultimate precision is limited by the asymtotic growth of
the perturbation theory series.
The coupling constant, though important, is still an
artificial parameter and the knowledge of its precise
numerical value does not suffice 
for computing observables with sufficiently high 
precision.

The investigation of the 
$\tau$ system can be performed in N$^4$LO without any free parameters
with the use of the internal charge $a$ 
(and even N$^5$LO with the single
free parameter $k_3$
which does not affect the conclusion about
the asymptotic structure of perturbation theory series).
However the $\MSsch$ scheme coupling can be expressed through $a$ only 
up to NNLO because of the unknown coefficient $k_3$. 
The extraction of
the $\overline{\rm MS}$ charge from $a$ can be made through the relation 
\be
\label{mstoarel}
 {\al_s^{\overline{\rm MS}}(M_\tau)\over \pi}=a - 1.64 a^2 + 15.7 a^3
+ (49.6 - k_3) a^4 +\ldots
\ee
with the reasonably fast convergence for $k_3 = 25$ or $k_3 = 100$.

\section{Infra-red 
fixed point as a model for an infinite perturbation theory series}
The accuracy of approximating a function with the sum of a finite
number of term of its 
asymptotic series 
depends strongly on 
the analytic structure of this function.
Generally, there is an infinite number of ways to sum an asymptotic 
series with quite different results. 
Therefore estimates of accuracy based on asymptotic series along 
can be rather misleading.
To discuss this issue in more detail
(though for illustrative purposes only)
we consider a model for the exact function as source for
the perturbative expansion (or recipe of the infinite
resummation).
The model uses the existence of the infra-red 
fixed point for the running coupling
in the third order of perturbation theory
which allows one to extrapolate running to the origin.
In this particular case we can compare perturbative expansions
with an exact answer. 
This example allows one to check the general
conclusions about the asymptotic structure and the divergence of the 
series even if this is done in a model dependent way. 
The effective $\beta$-function is given by 
\be
\label{effbeta}
\beta_{eff}(a)= -\frac{9}{4} a^2  - 4 a^3  + 25.7 a^4  
+ (409.5 - \frac{9}{2} k_3) a^5 + O(a^6) 
\ee
where the only free parameter is $k_3$ because the 
$\beta_3$ coefficient in the $\MSsch$ scheme is known \cite{beta4}.
The third order approximation of the $\beta$-function
eq.~(\ref{effbeta}) possesses
an infra-red 
fixed point with the value $a_f=a(0)=0.384$. 

The integration in eq.~(\ref{intmom}) can be explicitly performed
with the third order $\beta$-function from eq.~(\ref{effbeta}).
With the initial value $a=0.111$
one obtains 
$m_0^f=0.1605$, $m_1^f=0.130954$ to be compared with the results of 
eqs.~(\ref{momff},\ref{num0},\ref{num1}).
The naive estimate of the accuracy does not always work 
for this resummation recipe for all the moments.
Indeed,
having the explicit model at hand one can generate an arbitrary 
number of terms of the perturbative expansion.
For the zeroth moment the series diverges with the pattern
\be
\label{pat}
m_0^{fix} = 0.111 + 0.028 + 0.019 + 0.013+0.014
+ 0.018 + 0.029 + 0.053 + 0.114+\ldots
\ee
giving an ultimate accuracy of only about 10\%.
This accuracy is obtained keeping the smallest term.
The sum of the first four terms gives the best estimate 
\[
m_0^{fix,best} = 0.111 + 0.028 + 0.019 + 0.013 = 0.171\pm 0.013
\]
to be compared with the exact result $m_0^f=0.1605$.
The central value is a bit too high but still within the uncertainty
given by the last term.  

For $m_1^{fix}$ one also finds a divergent series but with a much
faster decrease of the first few terms. The pattern of ``convergence'' is
given by the following huge expression which we display to show how
complicated
things can become. One has 
\begin{eqnarray}
\label{m1f}
m_1^{fix}&=& 0.111 + 0.013861  + 0.006197  + 0.001054 
+0.000480 + 0.000088   \nonumber \\
&+&0.000053 + 0.000016+0.000015 + 0.000014 + 0.000019 + 0.000026\nonumber \\
&+&0.000042 + 0.000072
+0.000135 + 0.000268 + 0.000568 + 0.001277 \ \nonumber \\
&+&0.003 + 0.0076 + 0.01997 + 0.055 + \ldots
\end{eqnarray}
The best estimate 
is formally given by the sum of the first ten terms 
\be
\label{best1}
m_1^{fix,best}=0.132795\pm 0.000014
\ee
according to the formal prescription for the 
the evaluation of precision.
The exact result $m_1^f=0.130954$,
however, does not fall into the tiny interval given by the
error bars in eq.~(\ref{best1}).
Therefore the formal criterion of the accuracy is violated in
this case: the discrepancy 
$m_1^{fix,best}-m_1^f=0.00184$ is not controlled by the smallest 
term of the asymtpotic expansion eq.~(\ref{m1f}). 
Still this discrepancy is small
and the actual 
accuracy for the first moment given by the asymptotic expansion
eq.~(\ref{m1f}) is 1.3\%. This
suffices for practical purposes.
One can easily see the difference between these
two observables which reflects the different numerical magnitude 
of the infra-red contributions.
Note also that the numerical magnitude of the smallest term 
of the expansion eq.~(\ref{m1f}) is very sensitive to the value of the third 
coefficient of the effective $\beta$-function. 
For the decay rate we find 
$r_\tau^f=0.1946$ and $r_\tau^f(1)=0.1527$
to be compared with eq.~(\ref{rtau},\ref{rtau0111},\ref{theoac}) and 
eqs.~(\ref{rtau1},\ref{rtau10111}).

>From eq.~(\ref{effbeta}) one sees that an
infra-red fixed point exists also in fourth order 
of the effective $\beta$-function for any $k_3<95.9$.
For these numerical values
the effect of $k_3$ on the
exact moments within the model 
is rather weak.
The pattern of ``convergence''
for the decay rate is mainly determined by the contribution of
the zeroth order moment (or even by the mixed $\tilde m_{20}$
moment) and is very close to the expressions in
eqs.~(\ref{altmom}-\ref{pat}).

In this model there are ways of accelerating
the convergence with nice results but they definitely 
cannot be justified for use in the general case.
Still our conclusion about the achievable precision 
within finite order
perturbation theory in fifth order remains valid.

\section{Conclusions}
Using the standard estimate of the accuracy of an asymptotic
series we have found 
that the theoretical precision 
in the perturbative description of $\tau$-decay
is already limited by the
asymptotic growth of the coefficients in fifth order of perturbation theory. 
This is a scheme invariant
statement. 
The accuracy of perturbative expansions for a reasonably general
set of observables cannot be better than 5\% - 10\%.
Taking a stricter attitude we claim that the zeroth order moment is not
computable within perturbation theory. Any consistent 
description of $\tau$-decay
data
at fifth order of perturbation theory requires exclusion of 
the zeroth
order moment from the list of observables (or it should not constitute
a dominant contribution). 
At fifth order of perturbation theory 
and with the present numerical value of the coupling, the first 
two moments
of the spectral density are too different to be simultaneously treated
by perturbation theory with an accuracy better than 5\% - 10\%.
Therefore one has to go beyond 
finite order perturbation theory to compare these two observables
if one requires a theoretical accuracy that exceeds
present experimental accuracy.
This implies the use of 
some procedure of resummation. The
resummation
procedure is not
defined uniquely and the result depends on the prescription chosen
\cite{Pivtau,renRS,groote}.
Moreover,
if one resums the infinite number of perturbation theory terms
the condensates have no invariant meaning anymore and their numerical
values may change \cite{renRS}.
Therefore, improving the theoretical accuracy for this system
seems to require the creation of a new paradigm.

The extraction of $\al_s$ from the $\tau$-decay rate 
and its comparison with $\al_s$-values determined
from other experiments does not appear to be 
the best test of perturbation theory for the $\tau$ system. 
The crucial test of
the applicability of perturbation theory for the $\tau$ system
would be
the simultaneous calculation of two 
observables (moments) with an appropriate accuracy. 
If the set of moments includes the zeroth moment
then the ultimate accuracy of finite order perturbative expansions
has been already reached.

\subsection*{Acknowledgments}
F.~Krajewski thanks S.~Groote for correspondence
and sending a Compendium on Adler`s function.
The work is supported in part by the Volkswagen 
Foundation under contract No.~I/73611 and 
by the Russian Fund for Basic Research under contracts
Nos.~97-02-17065 and 99-01-00091. 
A.A.~Pivovarov is Alexander von Humboldt fellow.


\begin{thebibliography}{99}
\bibitem{exp}ALEPH collaboration, Z. Phys. C76(1997)15,
Eur.Phys.J. C4(1998)409,\\ 
CERN-EP/99-026.
\bibitem{PDG}Particle Data Group, Review of Particle Properties,
Eur.Phys.J. C3(1998)1. 
\bibitem{phys_report}K.G.~Chetyrkin, J.H.~K\"uhn and A.~Kwiatkowski, 
Phys.~Rep.\ { 277} (1996) 189   
\bibitem{eek2}S.G.Gorishny, A.L.Kataev and S.A.Larin,
Phys.Lett. B259(1991)144;\\
L.R.Surguladze and M.A.Samuel, Phys.Rev.Lett. 66(1991)560, 2416(E),\\
Phys.Rev. D44(1991)1602.
\bibitem{eek2c}K.G.~Chetyrkin,  Phys.Lett. B391(1997)402 
\bibitem{SchTra84} 
K.Schilcher, M.D. Tran, Phys.\ Rev.\ D 29 (1984) 570.
\bibitem{Bra88}
E.Braaten, Phys.\ Rev.\ Lett.\ 53 (1988) 1606.
\bibitem{Bra89} 
E.Braaten, Phys.\ Rev.\ D 39 (1989) 1458.
\bibitem{NarPic88}
S.Narison, A.Pich,  Phys.\  Lett.\ B 211 (1988) 183.
\bibitem{BraNarPic92} 
E.Braaten, S.Narison, A.Pich, Nucl.\ Phys.\ B 373 (1992) 581.
\bibitem{SVZ}M.A.Shifman, A.I.Vainshtein, V.I.Zakharov, Nucl.Phys. B147(1979)385
\bibitem{RG}N.N.Bogoliubov and D.V.Shirkov, Quantum fields (Benjamin, 1983).
\bibitem{cont}
C.Bernard, A.Duncan, J.LoSecco, S.Weinberg,
Phys.Rev. D12 (1975) 792;\\ 
E.Poggio, H.Quinn, S.Weinberg,
Phys.Rev. D13 (1976) 1958.
\bibitem{cont1}R.Shankar, Phys.Rev. D15 (1977) 755.
\bibitem{cont2}K.G.Chetyrkin, N.V.~Krasnikov, A.N.Tavkhelidze,
Phys.Lett. 76B (1978) 83.
\bibitem{Pivtau}A.A. Pivovarov, Sov. J. Nucl. Phys. 54(1991) 676;
Z.Phys. C53(1992)461; Nuovo Cim. 105A(1992)813.
\bibitem{effsch}
G. Grunberg, Phys.Lett. 95B(1980)70, Erratum-ibid. 110B(1982)501 
\bibitem{ksch}
N.V.Krasnikov, Nucl.Phys. B192 (1981) 497
\bibitem{kksch}A.L. Kataev, N.V.Krasnikov, A.A.Pivovarov,
Phys.Lett. 107B (1981) 115;\\ 
Nucl.Phys. B198 (1982) 508 
\bibitem{effDh}A.Dhar and V.Gupta, Phys.Rev. D29(1984)2822 
\bibitem{prl}S.Groote, J.G.K\"orner, A.A.Pivovarov, K.Schilcher,
Phys.Rev.Lett. 79 (1997) 2763
\bibitem{brodsky1}
S.J. Brodsky, J.R. Pelaez, N. Toumbas,
Phys.Rev.D60(1999)037501
\bibitem{brodsky}J.R. Pelaez, S.J. Brodsky, N. Toumbas, 
SLAC-PUB-8147, May 1999.\\
Talk given at 34th Rencontres de Moriond: 
QCD and Hadronic Interactions, \\
Les Arcs, France, 20-27 Mar 1999; hep-ph/9905435 
\bibitem{renRS}N.V.~Krasnikov, A.A.~Pivovarov, Mod.Phys.Lett. 
A11(1996)835.
\bibitem{DP} F. Le Diberder and A. Pich,
Phys. Lett. B286 (1992) 147; B289 (1992) 165.    
\bibitem{beta4}T. van Ritbergen, J.A.M.Vermaseren and S.A.Larin,  
Phys.Lett. B400(1997)379
\bibitem{groote}S. Groote, J.G. K\"orner, A.A. Pivovarov,\\
Phys. Lett. B407(1997)66,
Mod.Phys.Lett. A13 (1998) 637
\end{thebibliography}
\end{document}